\documentclass[12pt,a4paper]{article}
\usepackage{lipsum}
\usepackage{authblk}
\usepackage[top=2cm, bottom=2cm, left=2cm, right=2cm]{geometry}
\usepackage{fancyhdr}

\renewenvironment{abstract}{%
\hfill\begin{minipage}{0.95\textwidth}
\rule{\textwidth}{1pt}}
{\par\noindent\rule{\textwidth}{1pt}\end{minipage}}
\makeatletter
\renewcommand\@maketitle{%
\hfill
\begin{minipage}{0.95\textwidth}
\vskip 2em
\let\footnote\thanks 
{\LARGE \@title \par }
\vskip 1.5em
{\large \@author \par}
\end{minipage}
\vskip 1em \par
}
\makeatother
\usepackage[latin1]{inputenc}
\usepackage{longtable}
\usepackage[pdftex]{graphicx}
\usepackage{units}
\usepackage{amsmath}
\usepackage{amsfonts}
\usepackage{amssymb}
\usepackage{float}
\usepackage{lscape}
\usepackage{ifthen}
\usepackage{color}
\usepackage{booktabs}
\usepackage{textcomp}
\usepackage[section]{placeins}
\usepackage{minitoc}
\usepackage[numbers,sort&compress]{natbib}
\usepackage{subfigure}
\usepackage{epstopdf}
\usepackage{physics}
\usepackage{hyperref}
\usepackage{url} % Written by Donald Arseneau
\usepackage{etoolbox}
\makeatletter
\patchcmd{\chapter}{\if@openright\cleardoublepage\else\clearpage\fi}{}{}{}
\makeatother

\begin{document}


\title{Results from the ANTARES Neutrino Telescope}
\author[1,2]{M. Spurio\thanks{\large on behalf of the ANTARES Collaboration}}
\affil[1]{{Dipartimento di Fisica e Astronomia Universit\`{a} di Bologna}}
\affil[2]{INFN, Sezione di Bologna. \url{spurio@bo.infn.it}}

\maketitle

\begin{abstract}
A primary goal of a deep-sea neutrino telescopes as ANTARES is the search for astrophysical neutrinos in the TeV-PeV range. ANTARES is today the largest neutrino telescope in the Northern hemisphere. After the discovery of a cosmic neutrino diffuse flux by the IceCube, the understanding of its origin has become a key mission in high-energy astrophysics. ANTARES makes a valuable contribution for sources located in the Southern sky thanks to its excellent angular resolution in both the muon channel and the cascade channel (induced by all neutrino flavors). 
 Assuming various spectral indexes for the energy spectrum of neutrino emitters, the Southern sky and in particular central regions of our Galaxy are studied searching for point-like objects and for extended regions of emission. 
In parallel, by adopting a multimessenger approach, based on time and/or space coincidences with other cosmic probes, the sensitivity of such searches can be considerably augmented. 
ANTARES has participated to a high-energy neutrino follow-up of the gravitational wave signal GW150914, providing the first constraint on high-energy neutrino emission from a binary black hole coalescence. ANTARES has also performed indirect searches for Dark Matter, yielding limits for the spin-dependent WIMP-nucleon cross-section that improve upon those of current direct-detection experiments.
\end{abstract}

%\end{frontmatter}

% \linenumbers

\section{Introduction\label{sez:intro} }
%%%%%%%%%%%%%%%%%%%%%%%%%%%%%%%%%%%%%%%%%%%%%%%%%%%%%%%%%%%%%%
%%%%%%%%%%%%%%%%%%%%%%%%%%%%%%%%%%%%%%%%%%%%%%%%%%%%%%%%%%%%%%

The ANTARES neutrino telescope has been running in its final configuration since 2008; it comprise an array of 885 photomultipliers tubes housed in optical modules, detecting the Cherenkov light induced by charged particles produced by neutrino interactions in and around the instrumented volume \cite{ANTA18}. 
In this paper, the main recent results of ANTARES are summarized, focusing in particular on the quest for the origin of the IceCube astrophysical neutrino signal.

Astrophysical point-like sources of neutrinos can principally be individuated looking for an excess of muons from charged current (CC) interactions of $\nu_\mu$ in the proximity of the detector. 
The high rate of downgoing muons from the interactions of cosmic rays in the Earth's atmosphere usually restricts such searches to events with upwards going muons, or only a few degrees above the horizon. 
The primary background to such searches is due to atmospheric neutrinos and those few atmospheric muons mis-reconstructed as up-going. 
The long scattering length of light in seawater provides an excellent directional resolution on the $\nu_\mu$ of $\sim 0.4^\circ$ for an $E^{-2}$ source. 

Recently the collaboration has finalized an efficient, likelihood-based reconstruction method for cascade-like events. Cascades are mainly induced by neutral current (NC) interactions, and $\nu_{e}$ and $\nu_{\tau}$ CC interactions.
The effective area to cascade events is generally lower than to $\nu_{\mu}$ CC interactions, due to the very long range of the outgoing $\mu$. Additionally, the angular resolution for through-going muons is superior. However, in the cascade channel the energy deposited in the detector is more strongly correlated with the energy of the primary neutrino.  

\begin{figure}[tb]
\begin{center}
\includegraphics[width=0.7\textwidth]{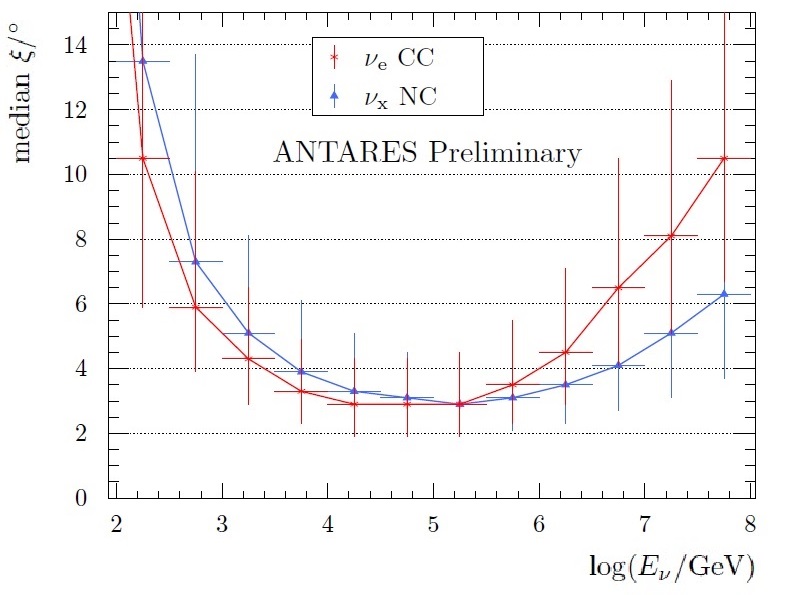}
\caption{Angular resolutions for Monte Carlo NC (blue) and CC (red) events reconstructed with the new shower algorithm \cite{ICRC07}. } 
\label{fig:sho}      
\end{center}
\end{figure}

The new algorithm allows to reconstruct the neutrino direction with a median angular resolution of about 3$^\circ$ in the $\sim 10$-$300$~TeV range, Fig. \ref{fig:sho}, and the deposited energy with a resolution of $\sim 5\%$, although the latter is limited by the total ANTARES systematic energy uncertainty of approximately $10\%$. 
Below $10$~TeV, the resolutions worsen due to a decreasing number of detected photons, while above $300$~TeV, the events begin to saturate the detector. 
The algorithm has been included in the standard reconstruction framework; it is the basis for the first analyses presented here using both tracks and cascades. 
%%%%%%%%%%%%%%%%%%%%%%%%%%%%%%%%%
%%%%%%%%%%%%%%%%%%%%%%%%%%%%%%%%%

%%%%%%%%%%%%%%%%%%%%%%%%%%%%%%%%%%%%%%%%%%%%%%%%%%%%%%%%%%%%%%
%%%%%%%%%%%%%%%%%%%%%%%%%%%%%%%%%%%%%%%%%%%%%%%%%%%%%%%%%%%%%%
\section{The neutrino sky in 2016\label{sez:physics.0}} 
%%%%%%%%%%%%%%%%%%%%%%%%%%%%%%%%%%%%%%%%%%%%%%%%%%%%%%%%%%%%%%
%%%%%%%%%%%%%%%%%%%%%%%%%%%%%%%%%%%%%%%%%%%%%%%%%%%%%%%%%%%%%%

The scientific landscape of high-energy neutrino astrophysics has significantly changed after the IceCube (IC) evidences for extraterrestrial high-energy neutrinos \cite{ICsci13,PRL14}
The identification of a cosmic neutrino flux up to PeV over the background of atmospheric neutrinos has been obtained by the IC collaboration using selection criteria in a restricted fiducial volume resulting in the so-called High Energy Starting Events (HESE) \cite{hese}.
The largest fraction of HESE are showers for which the angular determination is poor (10-20$^\circ$). 
The HESE flux is still compatible with flavor ratios $\nu_e:\nu_\mu:\nu_\tau= 1:1:1$, as expected from charged meson decays in CR accelerators and neutrino oscillation on their way to the Earth.

However, the recent measurement of the diffuse astrophysical muon neutrino flux using six years of IC data \cite{ICmu6} shows some tensions with the hypothesis of a symmetric Northern and Southern neutrino sky.
The IC measured flux of neutrino-induced upgoing muons shows an excess of events with respect to the purely atmospheric origin at the level of 5.6$\sigma$ in the neutrino energy range between 191 TeV and 8.3 PeV.
These up-going tracks are induced exclusively from $\nu_\mu$ from the Northern sky.
The excess corresponds to an astrophysical $\nu_\mu+\overline \nu_\mu$ flux $\Phi(E_\nu)=\Phi_0 E^{-\Gamma}$ with
\begin{equation}
\Phi(E_\nu) = 0.90^{+0.30}_{-0.27}\cdot (E_\nu/100 \textrm{TeV})^{2.13\pm 0.13 } 
\label{eq:1}
\end{equation}
in units of $10^{-18}$ GeV$^{-1}$ cm$^{-2}$ s$^{-1}$ sr$^{-1}$. 
When this flux from the Northern sky is compared with the combined analysis, with mainly cascade-like events largely originating from the Southern hemisphere, a tension at level of $3.3\sigma$ is present.

A simple explanation to this tension is that a galactic component is present in the signal, in addition to a completely diffuse one due to unresolved extragalactic sources \cite{spurio}. 
The extragalactic component is expected with harder spectral index ($\Gamma \simeq 2$) and smaller normalization factor with respect to that of the galactic component.
The softer spectral index of the galactic component ($\Gamma \simeq 2.5$) could be connected to neutrino production from interactions of cosmic rays close to the sources or during the propagation in the Galaxy. 

\begin{figure}[hbt]
\begin{center}
\includegraphics[width=0.80\textwidth]{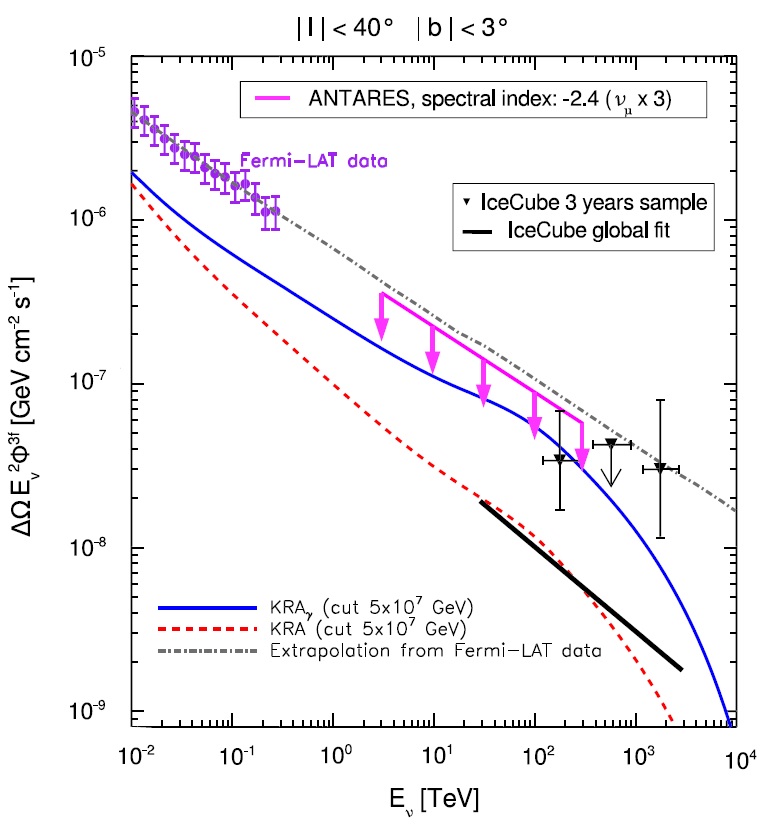}
\caption{\small  ANTARES 90\% C.L upper limit (magenta) for the search for an excess of events from the central Galactic region, compared to the expected neutrino flux (dot-dashed line) extrapolated from the \textit{Fermi}-LAT diffuse $\gamma$-ray flux up to IceCube energies and to the neutrino expectations as computed by Gaggero et al.} 
\label{fig:GC-fluxes}
\end{center}
\end{figure}

Cosmic rays in our Galaxy will collide with the interstellar medium producing pions and, hence, neutrinos. Direct evidence for these processes comes from observations by \emph{Fermi}-LAT  of the diffuse galactic $\gamma$-ray background \cite{LATdiff}. 
It is also interesting that the number of IC HESE in the $E>100$~TeV range with angular reconstructions consistent with the inner galactic plane corresponds to a flux consistent with that observed in $\gamma$ rays, as shown in Fig. \ref{fig:GC-fluxes}. 
The large uncertainty in the arrival directions of cascade-like HESE, and their low number, makes this comparison difficult. 
%%%%%%%%%%%%%%%%%%%%%%%%%%%%%%%%%%%%%%
%%%%%%%%%%%%%%%%%%%%%%%%%%%%%%%%%%%%%%

ANTARES northern latitude is ideally suited to study the expected neutrino flux from the inner galactic plane, and a search has been performed looking in the regions of galactic longitude $|l| < 40^{\circ}$ and latitude $|b| < 3^{\circ}$, as reported in \cite{ANTA52}. 
The study has used nine off-zones, and has found no excess in the on-zone region. 
The resulting limits are shown in Fig. \ref{fig:GC-fluxes}. In particular, the hypothesis of a 1--1 relation between the $\gamma$-ray and neutrino flux from the inner galactic plane is ruled out at $90$\% C.L., showing that ANTARES is already testing the multimessenger $\gamma$--$\nu$--CR paradigm in our Galaxy. 
The present limits cannot rule out more detailed simulations of galactic cosmic-ray propagation \cite{CRAg}.

%%%%%%%%%%%%%%%%%%%%%%%%%%%%%%%%%%%%%%%%%%%%%%%%%%%%%%%%%%%%%%
%%%%%%%%%%%%%%%%%%%%%%%%%%%%%%%%%%%%%%%%%%%%%%%%%%%%%%%%%%%%%%
\section{Point-like neutrino sources\label{sez:PS}} 
%%%%%%%%%%%%%%%%%%%%%%%%%%%%%%%%%%%%%%%%%%%%%%%%%%%%%%%%%%%%%%
%%%%%%%%%%%%%%%%%%%%%%%%%%%%%%%%%%%%%%%%%%%%%%%%%%%%%%%%%%%%%%
%%%%%%%%%%%%%%%%%%%%%%%%%%%%%%%%%%%%%%%%%%%%%%%%%%
\begin{figure}[htb]
\begin{center}
\includegraphics[width=0.7\textwidth]{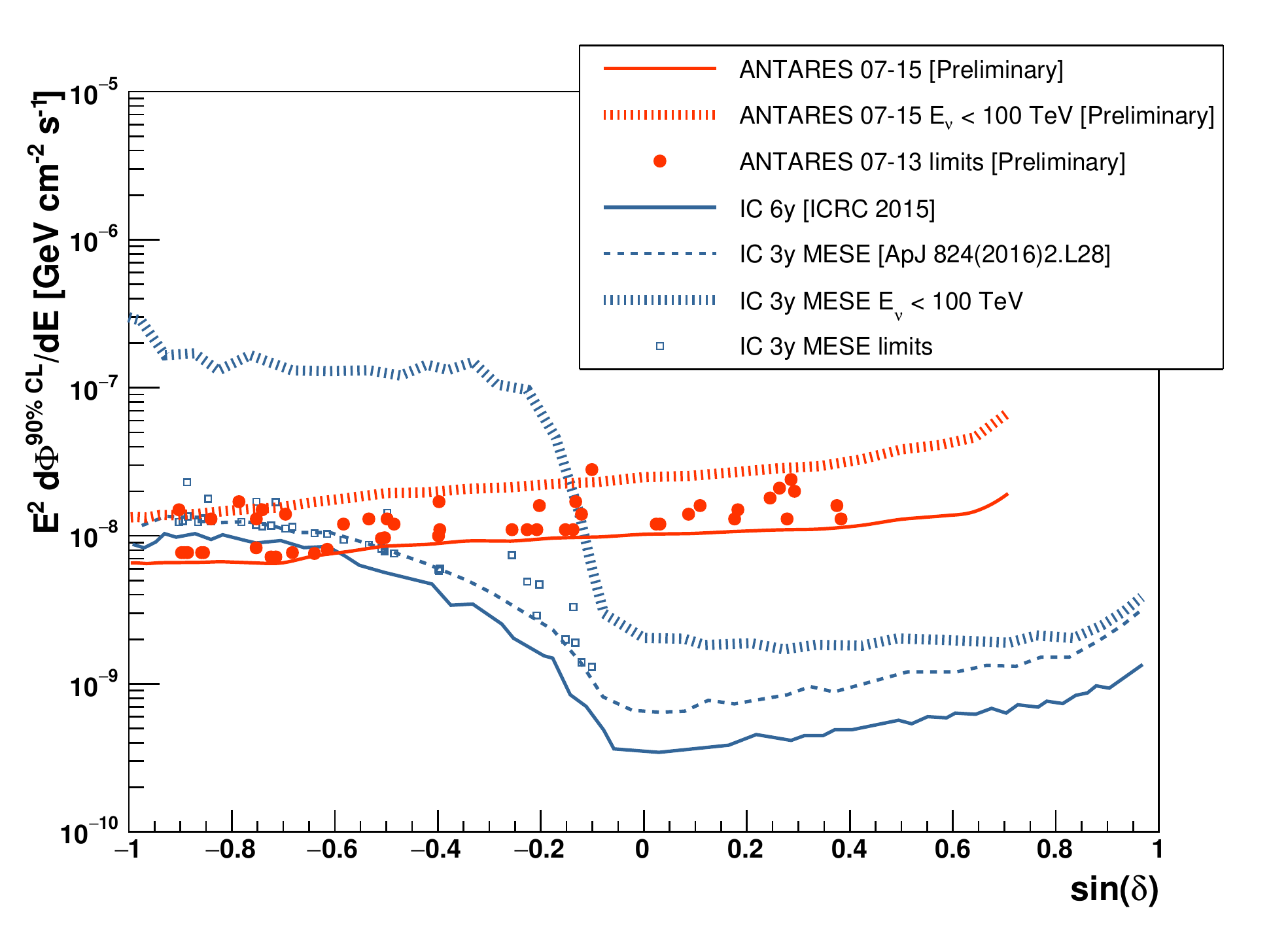} 
\caption{\small Sensitivities and limits for ANTARES (red) and IceCube (blue). Dots represent the limits. The sensitivities of ANTARES correspond to the all energy (continuous line) and $E_\nu < 100$ TeV (dashed line) sensitivities. The IceCube ones correspond to the 6y all-energy (continuous line), and  $E_\nu < 100$ TeV (think dashed line). For galactic sources, neutrino below 100 TeV are largely dominant.}
\label{fig:skymap}
\end{center}
\end{figure}
%%%%%%%%%%%%%%%%%%%%%%%%%%%%%%%%%%%%%%%%%%%%%%%%%%

The very good angular resolution for neutrino-induced muon events and the likelihood method used by ANTARES \cite{ANTA38} allows a strong suppression of both backgrounds in the signal direction, and a correspondingly good sensitivity to neutrino sources located in the Southern Hemisphere. 
A point source with a flavor-uniform flux and with $E^{-2}$ spectrum is expected to produce a cascade-to-track ratio of $3$:$10$ using the new cascade algorithm mentioned in \S \ref{sez:intro}.
The inclusion of the cascade channel has allowed a search for point-like sources using $1690$ days of effective livetime from $2007$ to $2013$ with a sensitivity at the level of $\sim 10^{-8}$ GeV$^{-1}$ cm$^{-2}$ s$^{-1}$ for $\delta < -40^{\circ}$.
After cuts, the sample consisted of $6490$ muon-track events, and $172$ cascade events, with an estimated contamination of $10$\% mis-reconstructed atmospheric muons in each.
An untargeted point-source search and a search over a list of pre-specified candidates have been applied to this data. Also the HESE have been included in the candidate list whereby for those search was extended to a few degrees around their direction depending on the respective uncertainty in their direction.  Also the galactic center was studied assuming different source extensions from 0 to 5 degrees. No significant excess has been observed. The resulting limits on point-like sources are given in Fig. \ref{fig:skymap}. 

\begin{figure}[htb]
\centering
\includegraphics[scale=0.6]{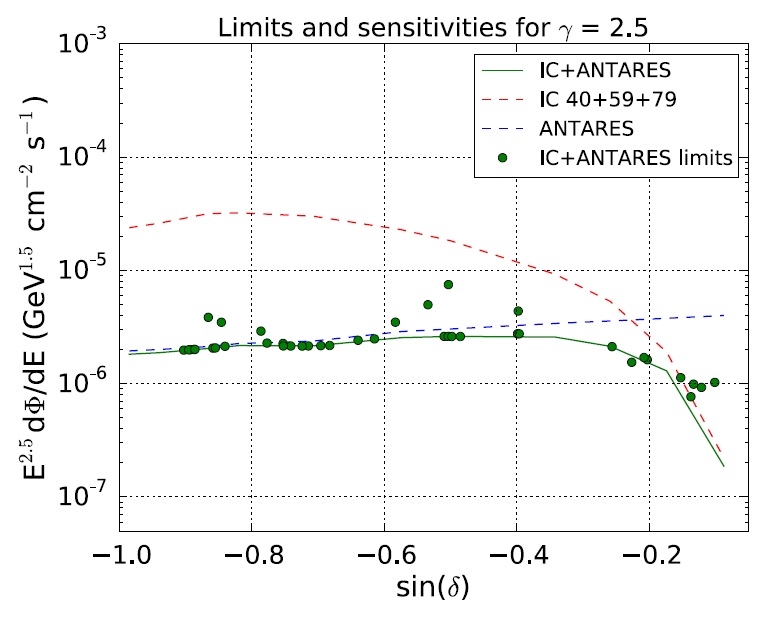}
\caption{\small Sensitivities (lines) and limits (dots) to an E$^{-2.5}$  flux, using ANTARES (blue), IceCube (red), and combined (green) data, as a function of $\sin$ of the declination $\delta$. }
\label{fig:AIC}       
\end{figure}

A joint ANTARES and IceCube search for a neutrino excess from selected sources in the Southern hemisphere is detailed in \cite{ANTA48}. 
The ANTARES contribution is dominant for declination $ <-15^\circ$. 
In fact, ANTARES is more sensitive to tracks produced by (relatively) low-energy $\nu_\mu$, while IceCube requires high-energy events to distinguish them from the huge background due to atmospheric muons. 
The overall sensitivity of each detector is a function also of the background rate, and of the angular and energy resolutions.
The results of the combined search are shown in Fig. \ref{fig:AIC}, for the case of an $E^{-2.5}$ source spectrum. 
No significant cluster is found and the combined analysis improves the limits set by each experiment. 
%%%%%%%%%%%%%%%%%%%%%%%%%%%%%%%%
%%%%%%%%%%%%%%%%%%%%%%%%%%%%%%%%

\begin{figure}[htb]
\centering
\includegraphics[width=0.6\textwidth]{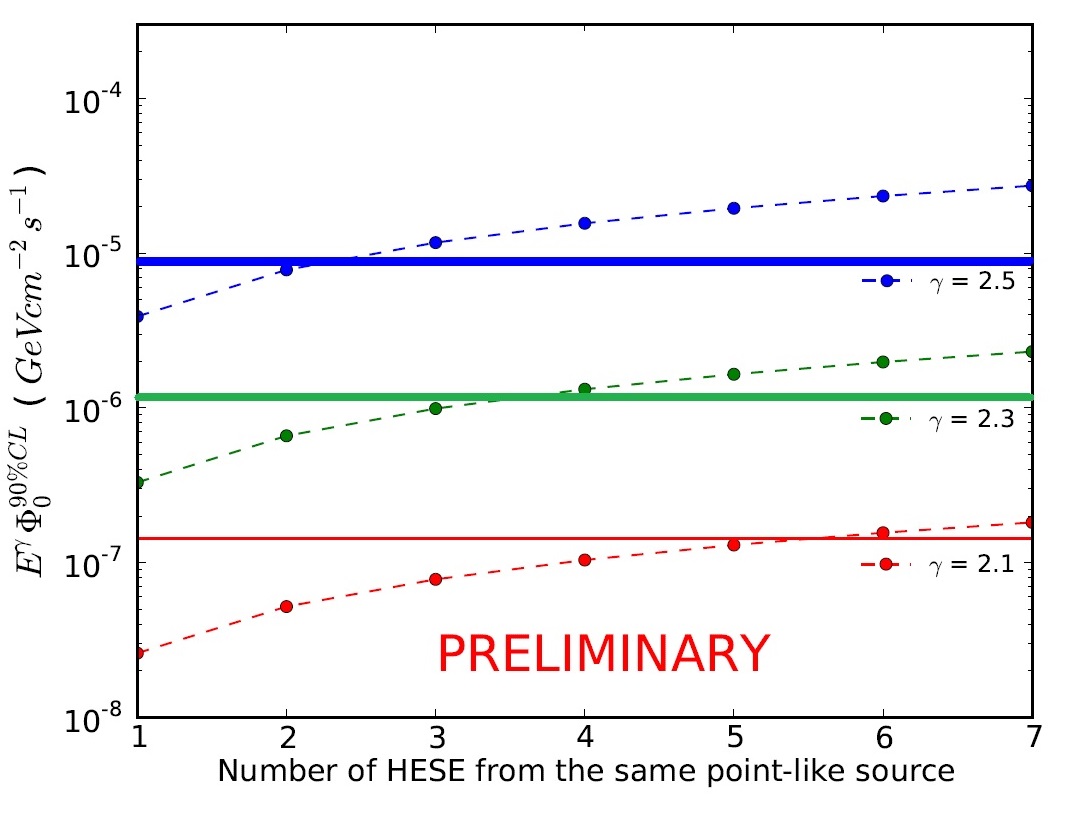}
\caption{\small ANTARES (2007-2012 data) upper limits (solid lines) at 90\% C.L.\ for different spectral indices $\gamma$, shown for a source at $\delta = -29^{\circ}$. These are compared to (dashed lines) the flux required to produce a given expected number of HESE. The range where the latter is greater than the former is excluded. } 
\label{fig:GC}       
\end{figure}

Because of the fact that the angular determination for most HESE is poor
(10-20$^\circ$), some authors \cite{gonza} argued that the cluster of events near the galactic plane could be due to a single, not identified source. 
ANTARES has searched for a possible excess in a wide region near the Galactic Center, without positive results. 
This limits the flux of such a source as a function of spectral index, shown by the solid lines in Fig. \ref{fig:GC} using data up to 2012. For instance, ANTARES rules out any single point source of neutrinos in the region of the Galactic Center with spectral index of $-2.5$ as having a flux corresponding to more than $2$~HESE. The flux limits is now significantly improved (publication in preparation) by including data up to 2015 and the shower channel.
%%%%%%%%%%%%%%%%%%%%%%%%%%%%%%%%%
%%%%%%%%%%%%%%%%%%%%%%%%%%%%%%%%%

The possibility that some HESE are coming from one (or several) transient sources in the Galactic Center has been considered in \cite{ICRC02}. This study has used a two-point correlation function and focuses on HESE located within 45$^\circ$ from the Galactic Center. This approach is sensitive to transient emission and requires neither prior on the burst timing structure nor on the electromagnetic emission.
Upper limits on the number of events and the duration of the flare have been computed.

%%%%%%%%%%%%%%%%%%%%%%%%%%%%%%%%%%%%%%%%%%%%%%%%%%%%%%%%%%%%%%
%%%%%%%%%%%%%%%%%%%%%%%%%%%%%%%%%%%%%%%%%%%%%%%%%%%%%%%%%%%%%%
\section{Diffuse flux and extended source \label{sez:physics.2}} 
%%%%%%%%%%%%%%%%%%%%%%%%%%%%%%%%%%%%%%%%%%%%%%%%%%%%%%%%%%%%%%
%%%%%%%%%%%%%%%%%%%%%%%%%%%%%%%%%%%%%%%%%%%%%%%%%%%%%%%%%%%%%%

In addition to the point-like candidate neutrino sources, several extended regions have been proposed as hadronic acceleration sites. 
ANTARES searched for an excess of the neutrino flux from these regions using 'on-zones' defined by specific templates, which are compared to 'off-zones' of the same size and shape, but offset in right ascension. As in the case of the Galactic Center region, the off-source zones give an unbiased estimate of the background in the source region in a way that is independent of simulations. 

The Fermi bubbles are giant regions of $\gamma$-ray emission extending out of the Galactic Center, and are proposed hadronic acceleration site, with neutrinos expected from $p$--$p$ collisions. A first search in ANTARES data from 2008--2011 for emission from these regions was presented in \cite{ANTA36}.
%%%%%%%%%%%%%%%%%%%%%%%%%%%%%%%%%
\begin{figure}[htb]
\centering
\includegraphics[width=1.0\textwidth]{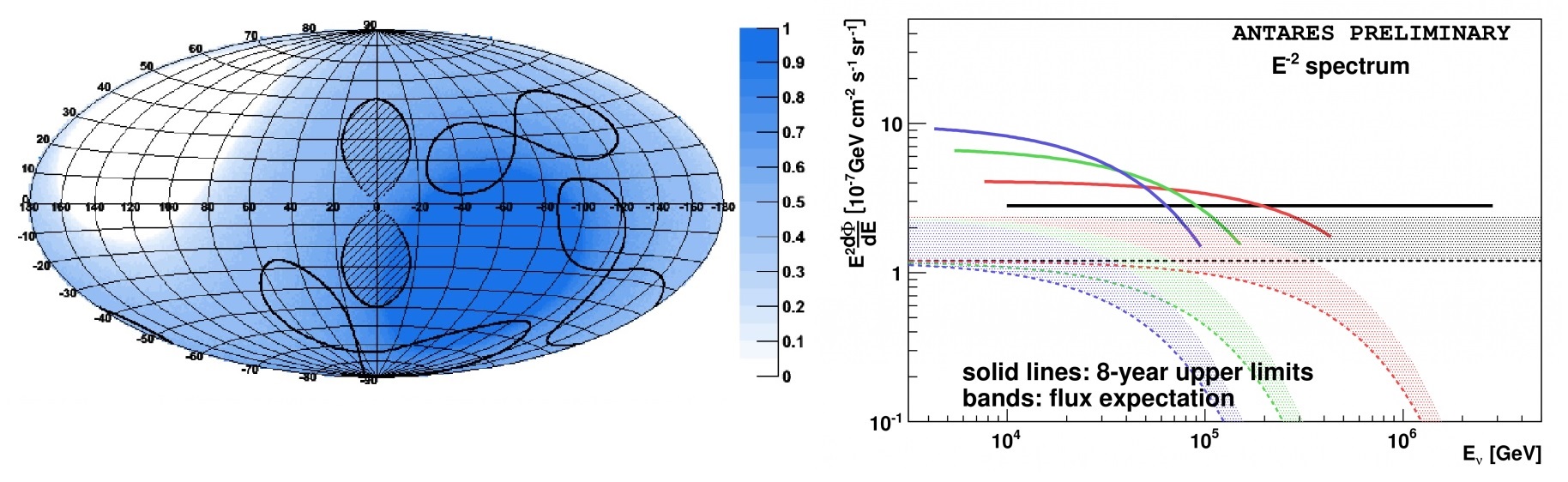}
\caption{\small (Left) On- and off-zone regions for the Fermi bubble search, compared to the ANTARES visibility (blue shading). (Right) $90$\% C.L. upper limits (lines) on the neutrino flux from the Fermi bubbles, compared to (shaded regions) expectations for different spectral shapes.}
\label{fig:fermi-bubbles}
\end{figure}
%%%%%%%%%%%%%%%%%%%%%%%%%%%%%%%%%

The on- and off-zone regions used in the Fermi bubble analysis are shown in Fig. \ref{fig:fermi-bubbles} (left). Flavor-uniform $E^{-2.0}$ and $E^{-2.18}$ neutrino fluxes have been assumed, where the latter is motivated by the best-fit proton spectrum \cite{luna}.
Exponential cut-offs at energies of $500$, $100$, and $50$ TeV have also been tested.
A slight excess has been determined in the source region using data from 2007 to 2015, corresponding to a $1.5~\sigma$ significance. The corresponding upper limits on an $E^{-2.18}$ neutrino flux are compared in Fig. \ref{fig:fermi-bubbles} (right) to the expectations.

%%% DIFFUSE
Connected with the expected level of the diffuse flux given in Eq. \ref{eq:1}, different strategies for the search of a diffuse cosmic neutrino flux have been adopted. The optimal method makes use of both muon tracks and cascade events \cite{icrc10}.
At present, the results are obtained with two independent analyses, searching for high energy neutrinos in the track and cascade channels separately.

The cascades reconstructed in the detector produced by NC and $\nu_e,\nu_\tau$ CC interactions \cite{ICRC07} are selected with a chain based on the topology of the hit distributions in the detector, allowing a good rejection of atmospheric muons. 
Thanks to the extremely good energy resolution in this channel, to suppress atmospheric neutrinos a cut on the energy estimator is defined using a blinded method. 
The best sensitivity is obtained for a cut at 30 TeV on the reconstructed energy.  
After the unblinding of data collected from 2007 to 2013, 7 events are observed in data when $5\pm 2$ are expected from atmospheric backgrounds. The cosmic signal observed by IceCube would produce $\sim 1.5$ events, depending on the spectral index.

\begin{figure}[tb]
\centering
\includegraphics[width=0.6\textwidth]{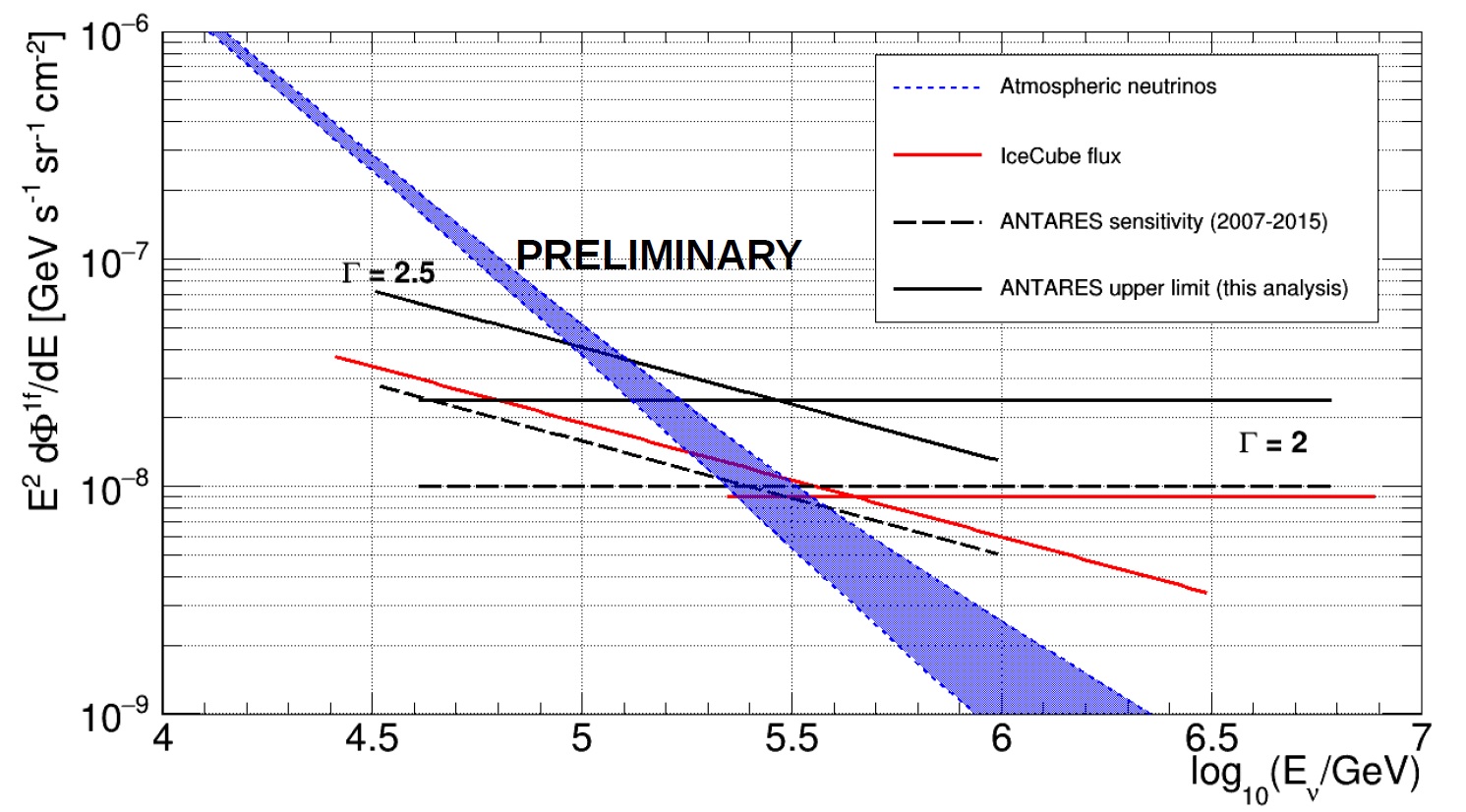}
\caption{\small 90\% C.L. upper limits for the combination of the track and shower analyses, compared to the sensitivity achievable with the entire ANTARES 2007-2015 data sample.} 
\label{fig:diffuse_limits}
\end{figure}

For the tracks induced by CC $\nu_\mu$, a blinded event selection chain based on the track quality parameters and on the number of selected hits is defined. This reduces the contamination from atmospheric muons below 0.5\%. An energy estimator based on an Artificial Neural Network provides discrimination between atmospheric and cosmic neutrinos. After the unblinding of data from 2007 to 2015 and the application of the energy cut, 19 events are observed while $13^{+3}_{-4}$ are expected from simulations of the atmospheric neutrinos. A cosmic flux analogous to that observed by IceCube would produce $\sim 3$ events.
The resulting limits on an $E^{-2}$ flux are given in Fig. \ref{fig:diffuse_limits}.
%%%%%%%%%%%%%%%%%%%%%%%%%%%%%%%%%%%%%%%%
%%%%%%%%%%%%%%%%%%%%%%%%%%%%%%%%%%%%%%%%

%%%%%%%%%%%%%%%%%%%%%%%%%%%%%%%%%%%%%%%%%%%%%%%%%%%%%%%%%%%%%%
%%%%%%%%%%%%%%%%%%%%%%%%%%%%%%%%%%%%%%%%%%%%%%%%%%%%%%%%%%%%%%
\section{Multimessenger astrophysics and transient phenomena \label{sez:multim}} 
%%%%%%%%%%%%%%%%%%%%%%%%%%%%%%%%%%%%%%%%%%%%%%%%%%%%%%%%%%%%%%
%%%%%%%%%%%%%%%%%%%%%%%%%%%%%%%%%%%%%%%%%%%%%%%%%%%%%%%%%%%%%%

Transient phenomena are particularly promising, since observations in the $\gamma$-ray spectrum (e.g. \textit{Fermi}-LAT experiment in the GeV and IACTs as MAGIC, VERITAS and H.E.S.S. in the TeV energy range) have shown that the high-energy universe is extremely variable.
In addition, restricting the searches to well-defined space-time windows decreases considerably the background, so that only one event may be enough to claim a discovery. 

The multimessenger effort of ANTARES to share data with other collaborations occurs in the framework of the TAToO program.
TAToO (Telescopes--ANTARES Target-of-Opportunity) performs near-real-time reconstruction of muons. 
If a sufficiently high-energy event, or an event with peculiar directions, is reconstructed as coming from below the horizon (i.e.\ the events most likely of astrophysical origin), an alert message is generated to trigger robotic optical telescopes (TAROT, MASTER, ZADKO) \cite{ANTA46} and radio telescopes (MWA) \cite{ANTA47}. The subsample with higher energy generates an alert also to the \emph{Swift}-XRT, the H.E.S.S. and the HAWC ground-based $\gamma$-ray detectors.
The very short alert-generation time (a few seconds) and half-sky simultaneous coverage of ANTARES makes it ideal for detecting transient signals, and optical and X-ray follow up observations have been initiated within 20~s and one hour respectively.

AGN have long been proposed as a source of high-energy cosmic rays and, hence, neutrinos. Blazars (AGN with jets pointed towards the Earth), exhibit bright flares which dominate the extragalactic $\gamma$-ray sky observed by \textit{Fermi}-LAT and IACTs.
Using multi-wavelength observations, several bright blazars have been reported by the TANAMI collaboration to lie within the $50$\% error bounds of the reconstructed arrival directions of the HESE PeV-scale events IC~14 and IC~20 observed by IceCube. 
ANTARES observes signal-like events from the two brightest blazars in the field of IC~20 \cite{ANTA42}, although this is also consistent with background fluctuations. 
A lack of such events from the field of IC 14 excludes a neutrino spectrum softer than $E^{-2.4}$ as being responsible for this event.
Later, the highest-energy `Big Bird' event (IC~35) has been observed during an extremely bright flare from the blazar PKS B1424-418, which lies within the $50$\% error region of the IC~35 arrival direction.
No ANTARES event has been detected from the direction of this source during the flaring period, constraining severely the mentioned model developed by the TANAMI collaboration.

Another analysis \cite{ANTA44} targets a sample of $41$ blazar flares observed by \textit{Fermi}-LAT and $7$ by the IACTs, searching for $\nu_\mu$ coming from the selected directions during flaring periods. 
The lowest pre-trial p-value of $3.3$\% has been found for the blazar 3C~279, which comes from the coincidence of one event with a flare occurred in 2008. However, the post-trial p-value is not significant. 

A similar method has been used to search for neutrino emission during the flares from galactic X-ray binaries. A total of $34$ X-ray and $\gamma$-ray-selected binaries have been studied, with no significant detections, allowing some of the more optimistic models for hadronic acceleration in these sources to be rejected at $90$\% C.L.

%% GRBs

Long-duration gamma-ray bursts (GRBs) have been proposed as a source of the highest-energy cosmic rays. ANTARES searched for a neutrino flux from GRBs considering two modelling for the emission processes: the description included in the NeuCosmA code \cite{neucosma}, and the `photospheric' model \cite{foto}.

\begin{figure}[htb]
\centering
\includegraphics[width=0.6\textwidth]{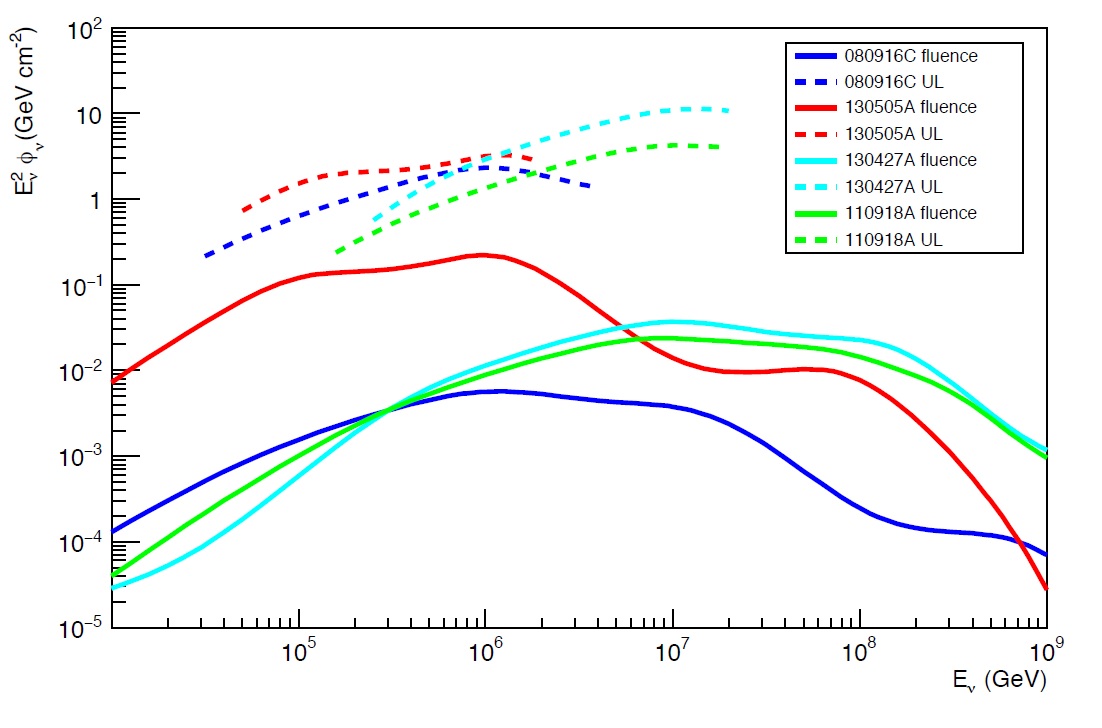}
\caption{\small Expected $\nu_\mu + \overline \nu_\mu$ fluence (solid line) and ANTARES 90\% C.L. upper limit (dashed line) on the selected GRBs, in the energy band where 90\% of the signal is expected to be detected by ANTARES (NeuCosmA model).}
\label{fig:grb}
\end{figure}

The main difference between the models is that the photospheric model predicts a neutrino flux that peaks at lower energies, due to fact that the emission occurs closer to the base of the jet.
In each case, the expected signal has been simulated on a burst-by-burst basis, and the detector response and background have been modelled using the exact sea conditions at the time of the burst. 
The ANTARES analysis using the NeuCosmA model has been applied to a sample of 296 bursts in \cite{ANTA33}, with no coincident neutrino events detected. 
Since then, one especially powerful burst GRB110918A, and the nearby burst GRB130427A, have been identified as promising candidates for neutrino detection. 
No coincident events have been observed from either bursts, with limits the neutrino fluences for those bright GRBs, as shown in Fig. \ref{fig:grb}. Similar results have been obtained for neutrinos from GRBs assuming production through the photospheric model. 
A very robust real-time analysis is also looking for GRB alerts distributed by the Gamma-ray Coordinates Network (GCN). This program is running since mid 2014 with more than 99\% efficiency.
%%%%%%%%%%%%%%%%%%%%%%%%%%%%%%%%%%%%%%%%%
%%%%%%%%%%%%%%%%%%%%%%%%%%%%%%%%%%%%%%%%%

A search for high-energy neutrino emission correlated with GRBs outside the electromagnetic prompt-emission time window has also been performed \cite{ANTAsta}. Using a stacking approach of the time delays between reported GRB alerts and spatially coincident muon-neutrino signatures, data recorded between 2007 and 2012 were analyzed. The respective timing profiles have been scanned for statistically significant accumulations within 40 days of the GRB, as expected from Lorentz Invariance Violation effects and some astrophysical models. No significant excess over the expected accidental coincidence rate could be found. The average strength of the neutrino signal is found to be fainter than one detectable neutrino signal per hundred GRBs at 90\% C.L.

Fast radio bursts (FRBs) are very energetic sources only seen in radio during few tens of milliseconds. The nature and the origin of such phenomena is still unknown. ANTARES has developed one on-line analysis for all events detected by the radio telescope PARKES in Australia (project SUPERB). Three alerts have been analyzed since November 2015. A similar analysis as the standard GRB one is also in progress to search for coincidences with past FRB detections.

A particular event occurred on September 1$^{st}$, 2015. The XRT onboard Swift followed an ANTARES high-energy neutrino, detecting an uncatalogued X-ray source at 8 arcmin from the neutrino direction \cite{GCN}.
This coincidence has generated the first telegram sent on behalf on the ANTARES Collaboration.
Different multi-wavelength observations (from IR, optic, radio, X-ray and TeV devices) have followed the alerts.
The X-ray follow-up of this source during five days has shown an X-ray flare of few days large. Optical data have indicated a bright star in the same location of the XRT source without significant variability. Additional multi-wavelength data have permitted to identify it as a young stellar object. There is therefore a very low probability that the X-ray source is associated with the observed neutrino. 
A paper on these multimessenger follow-up on this (likely fortuitous) coincidence is in preparation.

\section{ANTARES follow-up of the Gravitational Wave event detected by the A-LIGO interferometers\label{sez:gw}}
%%%%%%%%%%%%%%%%%%%%%%%%%%%%%%%%%%%%

Cataclysmic cosmic events can be plausible sources of both gravitational waves (GWs) and high-energy neutrinos (HEN). 
The merger of neutron stars and black holes, and potentially massive stellar core collapse with rapidly rotating cores, are expected to be significant sources of gravitational waves and potential emitters of high-energy neutrinos. The detection of such neutrinos in coincidence with a GW event would aid electromagnetic follow-up surveys by providing accurate source directions. Moreover, while HEN observations are probing the physics of relativistic outflows, GW are indicative of the dynamics and formation of the progenitor that drives the outflow.
Other possible sources include long and short GRBs but also low-luminosity or choked GRBs, with no or low $\gamma$-ray emissions.

Combining directional and timing information on HEN events and GW bursts through GW+HEN coincidences provides a novel way of constraining the processes at play in the sources. It also enables to improve the sensitivity of both channels relying on the independence of backgrounds in each experiment. 
The first search of that kind was performed with concomitant data from 2007 \cite{ANTA30}. 

A real breakthrough for multimessenger astrophysics occurred on September 14$^{th}$, 2015 at 9:50:45 UTC, when a GW event candidate was recorded by the LIGO Hanford (Washington, USA) and LIGO Livingston (Lousiana, USA) detectors, during their final Engineering Run ER8. The 90\% C.L. localization skymap of the potential source covers an area of roughly 590 degrees$^2$. According to preexisting MoU, the ANTARES and IceCube collaborations were informed, allowing the first HEN follow-up of a potentially significant gravitational wave detection. 
Three events have been found temporally coincident within $\pm$500 s of the GW by IceCube, while a background of two atmospheric neutrinos plus $\sim 2.2$ atmospheric muons are expected. ANTARES found no candidates, while 0.014 high-energy atmospheric neutrinos are expected in the same time windows. Thus, both searches are consistent with background. 

\begin{figure}[htb]
\centering
\includegraphics[width=0.6\textwidth]{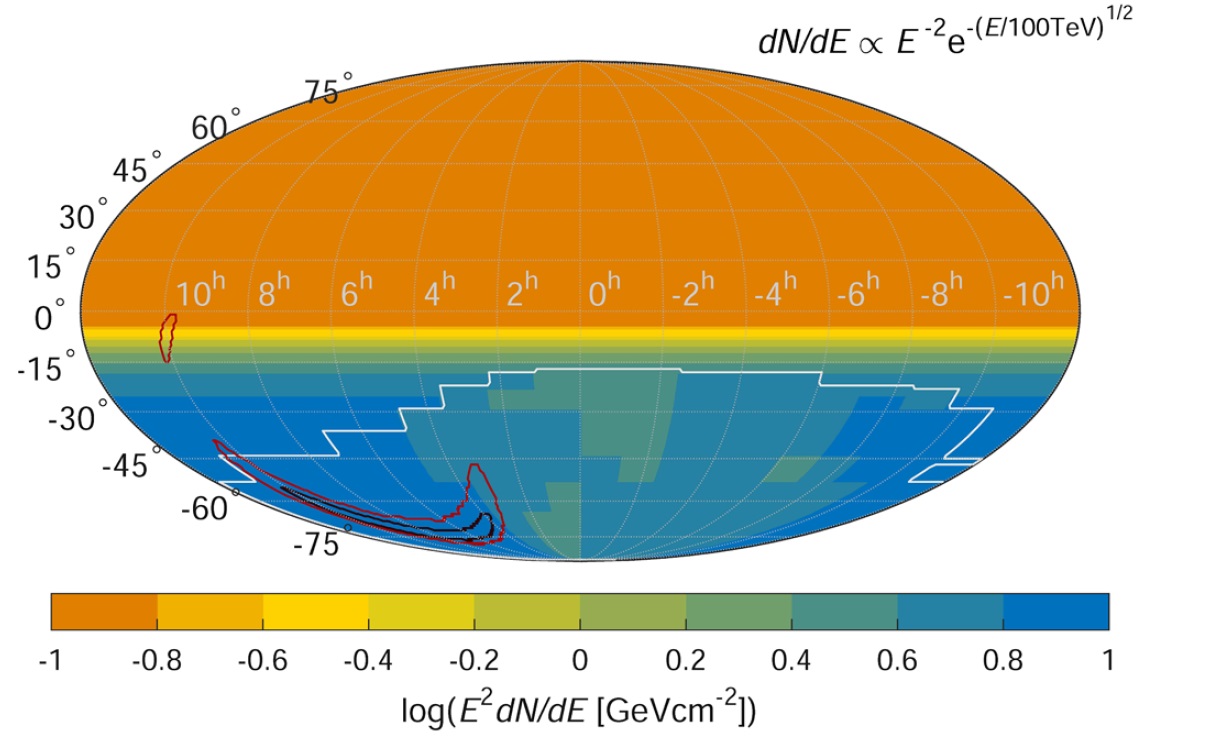}
\caption{\small Upper limit on the high-energy $\nu_\mu+\overline \nu_\mu$ spectral fluence from GW150914 as a function of source direction. The region surrounded by a white line shows the part of the sky in which ANTARES is more sensitive (close to nadir), while on the rest of the sky, IceCube is more sensitive. For comparison, the 50\% CL and 90\% CL contours of the GW sky map are also shown (red lines on the bottom left).} 
\label{fig:GW}
\end{figure}

The non-detection has been used to constrain neutrino emission from the gravitational wave transient event \cite{ANTA49}.
The obtained limit has been expressed in terms of total energy emitted in neutrinos, and computed both for a $E^{-2}$ generic spectrum and for a more realistic $E^{-2}$ spectrum with a cut-off at 100 TeV.
Figure \ref{fig:GW} shows the upper limits on the high-energy $\nu_\mu+\overline \nu_\mu$ spectral fluence $E^2 dN/dE$ as a function of source direction, for the spectrum with 100 TeV cut-off. 
ANTARES, mostly sensitive to 3 TeV-1 PeV neutrinos in the Southern hemisphere, constrains well the spectrum with a cut-off, where IceCube, sensitive to 200 TeV-100 PeV neutrinos in this region of the sky, is more constraining for the generic spectrum. This nicely illustrates the complementarity of the two telescopes.  
A similar analysis (with null result) has been performed for the second observed event, GW151226.
%%%%%%%%%%%%%%%%%%%%%%%%%%%%%%%%%%%%%%%%%
%%%%%%%%%%%%%%%%%%%%%%%%%%%%%%%%%%%%%%%%%

%%%%%%%%%%%%%%%%%%%%%%%%%%%%%%%%%%%%%%%%%%%%%%%%%%%%%%%%%%%%%%
%%%%%%%%%%%%%%%%%%%%%%%%%%%%%%%%%%%%%%%%%%%%%%%%%%%%%%%%%%%%%%
\section{Dark matter searches\label{sez:DM}} 
%%%%%%%%%%%%%%%%%%%%%%%%%%%%%%%%%%%%%%%%%%%%%%%%%%%%%%%%%%%%%%
%%%%%%%%%%%%%%%%%%%%%%%%%%%%%%%%%%%%%%%%%%%%%%%%%%%%%%%%%%%%%%

Neutrino telescopes can place limits on different WIMP dark-matter scenarios by limiting the neutrino flux expected from WIMP interactions in the Sun, Earth, Galactic Center, dwarf galaxies and galaxy clusters. Since the expected dark-matter density tends to be strongly peaked near the centers of these objects, and ANTARES has an excellent angular resolution, competitive limits can be set in the $E_{\rm WIMP} \gtrsim 50$~GeV range where the telescope is sensitive.
The geographical location of the detector is also an advantage compared to IceCube, since it allows a better visibility of the Galactic Center, being in the Northern hemisphere, and an observation of the Sun with less atmospheric background, being at intermediate latitude (and therefore observing the Sun less close to the horizon).

\begin{figure}[htb]
\centerline{\includegraphics[width=0.6\textwidth]{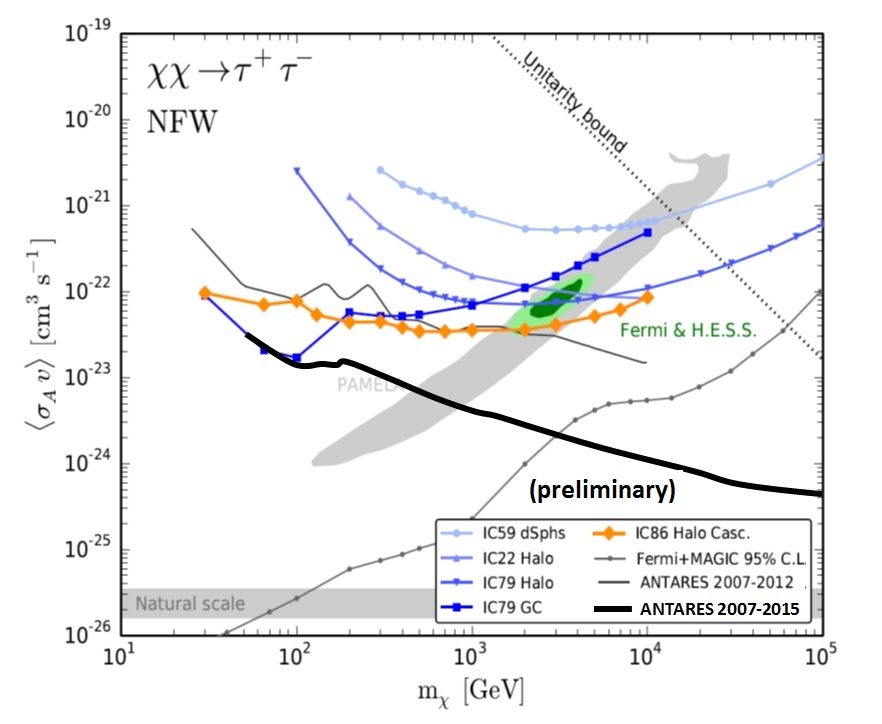} }
\caption{\small Limits on $\left< \sigma_A v \right>$ for the $\tau^+\tau^-$ channel from the Galactic Center as a function of the WIMP mass. The black line shows the ANTARES result using 2102 days of livetime and the same parameters of the NFW curve as IceCube. Plot modified from \cite{ICDm16}. } 
\label{fig:DM}
\end{figure}

The obtained results of the search for a neutrino flux from the direction of the Galactic Center with 2007-2015 data are the most competitive ones for neutrino telescopes (see the limits on the WIMP-WIMP velocity-averaged self-annihilation cross section $\sigma_A v$ shown in Fig. \ref{fig:DM}), given our better visibility of the Galactic Center compared to the South Pole \cite{ICDm16}. 
With respect to our previous paper on the same subject \cite{ANTA43}, the same parameters used by IceCube for the NFW dark matter density distribution is used in the computation of the new limit (paper in preparation).
%%%%%%%%%%%%%%%%%
%%%%%%%%%%%%%%%%%

\begin{figure}[htb]
\centerline{\includegraphics[width=0.6\textwidth]{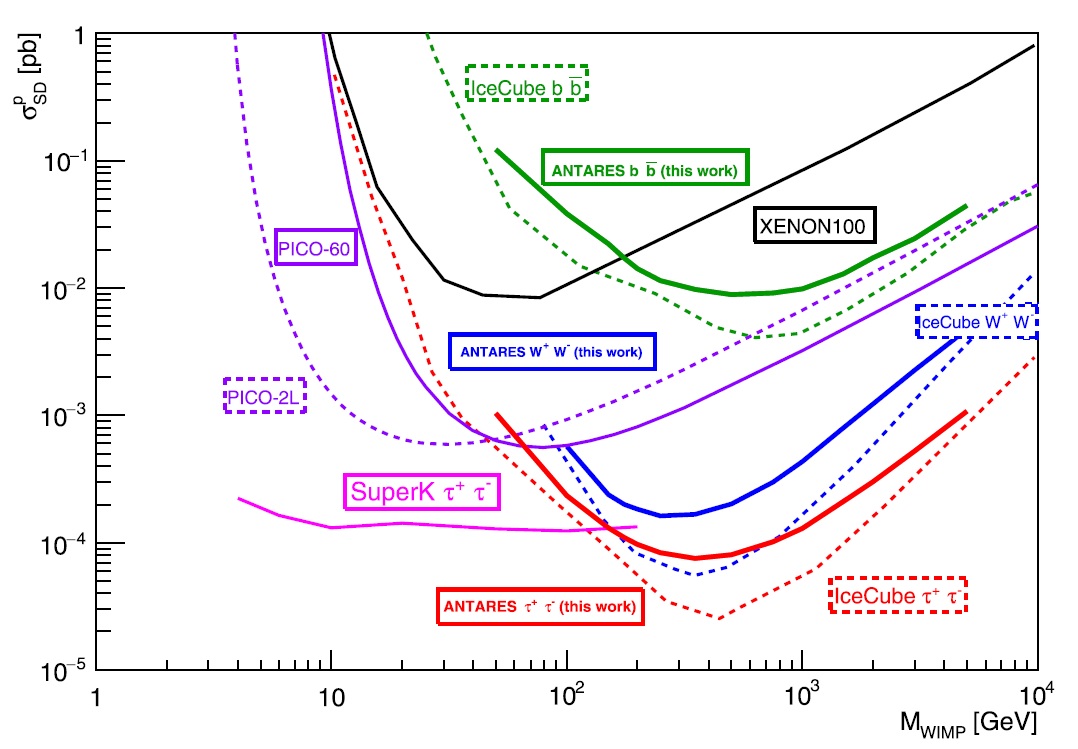} }
\caption{\small Limits from the Sun on the spin-dependent $\sigma^p_{\rm SD}$ WIMP-nucleon scattering cross-section as a function of WIMP mass for the $b^+b^-, \tau \bar{\tau}, W^+W^-$ channels. } 
\label{fig:DMsun}
\end{figure}

A new unbinned method has been implemented and applied on the analysis of events from the Sun direction. Fig. \ref{fig:DMsun} shows the limits on the spin-dependent (WIMP-proton) interaction cross section $\sigma^p_{\rm SD}$ from the observations of the Sun using different annihilation channels \cite{ANTA51}.
Neutrino telescopes produce the more stringent limits in the $M_{\rm WIMP} > 200$~GeV range, surpassing even the direct-detection experiments.
%%%%%%%%%%%%%%%%%
%%%%%%%%%%%%%%%%%

%%%%%%%%%%%%%%%%%%%%%%%%%%%%%%%%%%%%%%%%%%%%%%%%%%%%%%%%%%%%%%
%%%%%%%%%%%%%%%%%%%%%%%%%%%%%%%%%%%%%%%%%%%%%%%%%%%%%%%%%%%%%%
\section{Conclusions\label{sez:conclusions} }
%%%%%%%%%%%%%%%%%%%%%%%%%%%%%%%%%%%%%%%%%%%%%%%%%%%%%%%%%%%%%%
%%%%%%%%%%%%%%%%%%%%%%%%%%%%%%%%%%%%%%%%%%%%%%%%%%%%%%%%%%%%%%

The ANTARES telescope has demonstrated the great potential of deep-sea neutrino observatories in the Northern hemisphere. 
The scientific framework of neutrino astronomy has significantly changed after the IceCube detection of cosmic neutrinos, but without the identification of possible sources. 
ANTARES has an excellent angular resolution on both muon-track and cascade events, facilitated by the optical properties of deep-sea water. 
A new era in neutrino astronomy will begin in 2017, with the decommissioning of ANTARES, and the on-going construction of KM3NeT Phase 1, with a unique design of multi-PMT optical modules. 
When completed, Phase-1 will have an instrumented volume a factor of $\sim$ 3 larger than that of ANTARES and will pave the way to the multi-km$^3$ detector in the Mediterranean Sea able to monitor the Southern sky \cite{km3loi} with the sensitivity necessary for discoveries. 



%% References

%\nocite{*}
%\bibliographystyle{elsarticle-num}
%\bibliography{martin}



\end{document}